\documentclass{article}
\usepackage{amsmath,amssymb}
\usepackage{array}
\usepackage{amsfonts}
\usepackage{amsmath}
\usepackage{rotate,epsfig}
\usepackage{amsfonts}
\usepackage{lscape}
\usepackage{graphicx}
\usepackage{color}

\begin{document}

\title{Goodness-of-fit testing for the Cauchy distribution with application to financial modeling}
\author{ {{ M. Mahdizadeh}}${}^{\rm a,}$\footnote{Corresponding author.\newline {\it E-mail addresses:}
mahdizadeh.m@live.com (M. Mahdizadeh), e.zamanzade@sci.ui.ac.ir; ehsanzamanzadeh@yahoo.com (Ehsan Zamanzade)} \,and { Ehsan Zamanzade}${}^{\rm b}$  {\vspace{2mm}}\\
{\normalsize {\it $~^{\rm a}$Department of Statistics, Hakim Sabzevari University,  P.O. Box 397, Sabzevar, Iran}}\vspace{-0.1cm}\\
{\normalsize {\it $~^{\rm b}$Department of Statistics, University of Isfahan, Isfahan, 81746-73441, Iran}}\vspace{-0.1cm}}

\date{}
\maketitle

\thispagestyle{plain}
\setcounter{page}{1}

\begin{abstract}
\noindent This article deals with goodness-of-fit test for the Cauchy distribution. Some tests based on Kullback-Leibler information are proposed, and shown to be consistent. Monte Carlo evidence indicates that the tests have satisfactory performances against symmetric alternatives. An empirical application to quantitative finance is provided.
\end{abstract}

\section{Introduction}
A Cauchy random variable with location parameter $\mu \in \mathbb{R}$ and scale parameter $\sigma>0$, denoted by $X\sim C(\mu,\sigma)$, has probability density function
\begin{equation}\label{dens}
f_0(x;\mu,\sigma)=\left(\pi \sigma \left[ 1+\left(\frac{x-\mu}{\sigma}\right)^2 \right]\right)^{-1},\quad x\in \mathbb{R}.
\end{equation}
and cumulative distribution function
\begin{equation}\label{dist}
F_0(x;\mu,\sigma)=\frac{1}{2}+\frac{1}{\pi} \tan^{-1}\left(\frac{x-\mu}{\sigma}\right),\quad x\in \mathbb{R}.
\end{equation}

Sim\'{e}on Denis Poisson discovered the Cauchy distribution in 1824, long before its first mention by Augustin-Louis Cauchy. Early interest in the distribution focused on its value as a counter-example which demonstrated the need for regularity conditions in order to prove important limit theorems (see Stigler, 1974). Thanks to this special nature, the Cauchy distribution is sometimes considered as a pathological case. However, it can be used as a model for describing a wealth of phenomena. This is exemplified in the sequel.

This probability law describes the energy spectrum of an excited state of an atom or molecule, as well as an elementary particle resonant state. It can be shown quantum mechanically that whenever one has a state which decays exponentially with time, the energy width of the state is described by the Cauchy distribution (Roe, 1992). Winterton et al. (1992) showed that the source of fluctuations in contact window dimensions is variation in contact resistivity, and the contact resistivity is distributed as a Cauchy random variable. Kagan (1992) pointed out that the Cauchy distribution describes the distribution of hypocenters on focal spheres of earthquakes. An application of this distribution to study the polar and non-polar liquids in porous glasses is given by Stapf et al. (1996). Min et al. (1996)  found that Cauchy distribution describes the distribution of velocity differences induced by different vortex elements. An example in the context of  quantitative finance is provided in Section 4.

Many statistical procedures, employed in the above mentioned applications, assume that the random mechanism generating the data follows the Cauchy distribution. A parametric procedure usually hinges on the assumption of a particular distribution. It is, therefore, of utmost importance to assess the validity of the assumed distribution. This is accomplished by performing a goodness-of-fit test. In this article, we suggest some tests of fit for the Cauchy distribution. They are modifications of a test based on Kullback-Leibler (KL) information criterion, previously studied by the authors. A large body of literature has grown around developing goodness-of-fit tests using the information-theoretic measures such as the KL distance. This approach has been successfully applied for many distributions, including normal, uniform, exponential, inverse Gaussian and Laplace, among others. See for example Vasicek (1976), Dudewicz and van der Meulen (1981), Grzegorzewski and Wieczorkowski (1999), Mudholkar and Tian (2002), and Choi and Kim (2006).

Section 2 is given to a review of the existing tests. The new goodness-of-fit tests are presented in Section 3. Power properties of these tests are assessed by means of Monte Carlo simulations. The results are reported in Section 4. To illustrate the suggested procedures, a real data set is analyzed in Section 5. We end in Section 6 with a summary.

\section{Goodness-of-fit tests}
Given a random sample $X_1,\ldots,X_n$ from a population having a continuous density function $f(x)$, consider the problem of testing $H_0:f(x)=f_0(x;\mu,\sigma)$ for some $\mu \in \mathbb{R}$ and $\sigma>0$, where $f_0(x;\mu,\sigma)$ is given (\ref{dens}). The alternative hypothesis is $H_1:f(x)\neq f_0(x;\mu,\sigma)$ for any $\mu \in \mathbb{R}$ and $\sigma>0$.

The Cauchy distribution is a peculiar distribution due to its heavy tail and the difficulty of estimating its parameters (see Johnson et al., 1994). First, the method of moment estimation fails since the mean and variance of the Cauchy distribution do not exist. Second, the maximum likelihood estimates of the parameters are very complex. We therefore estimate $\mu$  and $\sigma$ by the median and the half-interquartile range which are attractive estimators because of their simplicity. Suppose $X_{(1)}\leq \cdots \leq X_{(n)}$ are the sample order statistics, and $\xi_p \,(0<p<1)$ is the sample $p$\,th quantile. Then, the two estimators are
given by
\begin{equation}\label{mhat}
\hat{\mu}= \left\{
\begin{array}{cc}
\frac{1}{2}(X_{(n/2)}+X_{(n/2+1)}) & \text{if $n$ is even}\\
X_{((n+1)/2)} & \text{Otherwise}
\end{array} \right.
\end{equation}
and
\begin{equation}\label{shat}
\hat{\sigma}=\frac{1}{2}(\xi_{0.75}-\xi_{0.25}).
\end{equation}

Suppose $F_0(x;\mu,\sigma)$ is defined as in (\ref{dist}). The best-known statistic for tests of fit is that of Kolmogorov-Smirnov given by
\begin{equation}\label{ks}
\textrm{KS}=\max_{i=1,\ldots,n} \left[\max\left\{\frac{i}{n}-F_0(X_{(i)};\hat{\mu},\hat{\sigma}),F_0(X_{(i)};\hat{\mu},\hat{\sigma})-\frac{i-1}{n} \right\} \right].
\end{equation}
Another powerful test, especially for small sample sizes, is based on the Anderson-Darling statistic defined as
\begin{equation}\label{ad}
A^2=-\frac{2}{n} \sum_{i=1}^n \Big[ (i-0.5)\log\Big\{F_0(X_{(i)};\hat{\mu},\hat{\sigma})\Big\}+(n-i+0.5)\log\Big\{1-F_0(X_{(i)};\hat{\mu},\hat{\sigma})\Big\} \Big]-n.
\end{equation}
The famous Cram\'{e}r-von Mises statistic,
\begin{equation}\label{cm}
W^2=\sum_{i=1}^n \left(F_0(X_{(i)};\hat{\mu},\hat{\sigma})-\frac{i-0.5}{n}\right)^2+\frac{1}{12 n},
\end{equation}
leads to an important goodness-of-fit test. The above three test are based on weighted distance between true and empirical distribution functions. G\"{u}rtler and Henze (2000) proposed a test based on the empirical characteristic function
$$\Psi_n(t)=\frac{1}{n} \sum_{j=1}^n \exp(itY_j)$$
of the standardized data $Y_j=(X_j-\hat{\mu})/\hat{\sigma}$, $j=1,\ldots,n$. The test statistic,
$$D_{n,\lambda}=n \int_{-\infty}^\infty \left| \Psi_n(t)-e^{-|t|} \right|^2 e^{-\lambda |t|}  \textrm{d}t,$$
is the weighted $L^2$ distance between $\Psi_n$ and the characteristic function of the standard Cauchy distribution, where $\lambda$ denotes a fixed positive weighting parameter. Large values of $D_{n,\lambda}$ imply rejection of $H_0$. After some algebra, an alternative representation of $D_{n,\lambda}$ is derived as
\begin{equation}\label{dn}
D_{n,\lambda}=\frac{2}{n} \sum_{j=1}^n \sum_{k=1}^n  \frac{\lambda}{\lambda^2+(Y_j-Y_k)^2}-4 \sum_{j=1}^n \frac{1+\lambda}{(1+\lambda)^2+Y_j^2}+\frac{2n}{2+\lambda}.
\end{equation}
{\bf Remark 1:} In practice, we use $\lambda=5$ which leads to a powerful test according to the simulation results reported by G\"{u}rtler and Henze (2000).

Recently, Mahdizadeh and Zamanzade (2016) proposed four new tests of fit for the Cauchy distribution. The first three of them are modifications of the tests introduced by Zhang (2002). The corresponding test statistics are
\begin{equation}\label{zk}
Z_K=\max_{i=1,\ldots,n} \left[(i-0.5)\log\left\{\frac{i-0.5}{n F_0(X_{(i)};\hat{\mu},\hat{\sigma})}\right\}+(n-i+0.5)\log\left\{\frac{n-i+0.5}{n (1- F_0(X_{(i)};\hat{\mu},\hat{\sigma}))}\right\}\right],
\end{equation}
\begin{equation}\label{za}
Z_A=-\sum_{i=1}^n \left[\frac{\log\Big\{F_0(X_{(i)};\hat{\mu},\hat{\sigma})\Big\}}{n-i+0.5}+\frac{\log\Big\{1-F_0(X_{(i)};\hat{\mu},\hat{\sigma})\Big\}}{i-0.5}\right]
\end{equation}
and
\begin{equation}\label{zc}
Z_C=\sum_{i=1}^n \left[\log\left\{ \frac{1/F_0(X_{(i)};\hat{\mu},\hat{\sigma})-1}{(n-0.5)/(i-0.75)-1}\right\}\right]^2.
\end{equation}
The fourth test utilizes the KL distance (see Kullback (1997)) between $f$ and $f_0$ given by
\begin{eqnarray}\label{kl dist}
D(f,f_0;\mu,\sigma)&=&\int_{-\infty}^{\infty} f(x) \log\left(\frac{f(x)}{f_0(x;\mu,\sigma)}\right) \textrm{d}x \nonumber \\
&=& -H(f)-\int_{-\infty}^{\infty} f(x) \log\left(f_0(x;\mu,\sigma) \right) \textrm{d}x,
\end{eqnarray}
where $H(f)$ is Shannon's entropy of $f$ defined as
$$H(f)=-\int_{-\infty}^{\infty} f(x) \log\left(f(x) \right) \textrm{d}x.$$
It is well known that $D(f,f_0;\mu,\sigma) \geq 0$ and the equality holds if and only if $f(x)=f_0(x;\mu,\sigma)$, almost surely. Therefore, $D(f,f_0;\mu,\sigma)$ can be regarded as a measure of disparity between $f$ and $f_0$.

Constructing a test based on (\ref{kl dist}) entails estimating the unknown quantities. The non-parametric estimation of $H(f)$ has been studied by many authors. Vasicek (1976) introduced a simple estimator which has been widely used in developing tests of fit. His estimator is
given by
\begin{equation}\label{V.est}
HV_{m,n}=\frac{1}{n}\sum_{i=1}^n \log\left\{\frac{n}{2m}\left(X_{(i+m)}-X_{(i-m)}\right)\right\},
\end{equation}
where $m$ (called window size) is a positive integer less than or equal to $n/2$, $X_{(1)}\leq \cdots \leq X_{(n)}$ are order statistics based on a random sample of size $n$, $X_{(i)}=X_{(1)}$ for $i<1$, and $X_{(i)}=X_{(n)}$ for $i>n$. Vasicek (1976) showed that (\ref{V.est}) is a consistent estimator of the population entropy. In particular, $HV_{m,n} \stackrel{p}{\rightarrow} H(f)$ as $m \rightarrow \infty$, $n \rightarrow \infty$ and $m/n \rightarrow 0$, where $\stackrel{p}{\rightarrow}$ denotes convergence in probability. Also,
$$\int_{-\infty}^{\infty} f(x) \log\left(f_0(x;\mu,\sigma) \right) \textrm{d}x$$
can be estimated by
\begin{equation}\label{I.est}
\frac{1}{n} \sum_{i=1}^n \log (f_0(X_i;\hat{\mu},\hat{\sigma})),
\end{equation}
which is consistent by virtue of law of large numbers. Mahdizadeh and Zamanzade (2016) suggested to use
\begin{equation}\label{Dinf}
\hat{D}_1=\exp\left\{-HV_{m,n}-\frac{1}{n} \sum_{i=1}^n \log (f_0(X_i;\hat{\mu},\hat{\sigma}))\right\}
\end{equation}
as the final test statistic. Large values of $\hat{D}_1$ provide evidence against the null hypothesis.

It is difficult to derive the null distributions of (\ref{ks})-(\ref{zc}) and (\ref{Dinf}) analytically. Monte Carlo simulations were then employed to determine critical values of a generic test statistic, say $T$. To this end, 50,000 samples were generated from $C(0,1)$ for each sample size $n=10,20,30,50$. The estimators (\ref{mhat}) and (\ref{shat}) were computed from any sample, and plugged into $T$. Finally, $1-\alpha$ quantile of the resulting values was determined which will be denoted by $\mathcal{T}_{1-\alpha}$. The composite null hypothesis is rejected at level $\alpha$ if the observed value of $T$ exceeds $\mathcal{T}_{1-\alpha}$.

\section{New tests}
In this section, we introduce some new testing procedures for the Cauchy distribution through altering (\ref{Dinf}). This is done by using other nonparametric entropy estimators which are set out here. Bowman(1992) studied the estimator
\begin{equation}\label{bow}
HB_n=-\frac{1}{n} \sum_{i=1}^n \log\left\{\hat{f}(X_i)\right\},
\end{equation}
where
$$\hat{f}\left(x\right)=\frac{1}{nh}\sum_{j=1}^n K\left(\frac{x-X_j}{h} \right),$$
and $K(.)$ is a symmetric kernel function which is chosen to be the standard normal density function. The
bandwidth $h$ is selected based on the normal optimal smoothing formula, $h=1.06 \,s\, n^{-1/5}$, where $s$ is the sample standard deviation.

Van Es (1992) considered estimation of functionals of a probability density and entropy in particular. He proposed the following estimator
\begin{equation}\label{van es}
HVE_{m,n}=\frac{1}{n-m}\sum_{i=1}^{n-m} \log\left\{\frac{n+1}{m}\left(X_{(i+m)}-X_{(i)}\right)\right\}+\sum_{i=m}^n \frac{1}{i} -\log\left\{\frac{n+1}{m}\right\},
\end{equation}
where $m$ is a positive integer less than $n$.

Ebrahimi et al. (1994) suggested two improved entropy estimators. The first one is equal to that of Vasicek plus a constant. This implies that the test based on this estimator is equivalent to $\hat{D}_1$. So it is not included in this study. The second estimator is given by
\begin{equation}\label{he2}
HE_{m,n}=\frac{1}{n}\sum_{i=1}^n \log\left\{\frac{n}{d_i m}(Y_{(i+m)}-Y_{(i-m)})\right\},
\end{equation}
where
$$
d_i=\left \{\begin{array}{lc}
1+\frac{i+1}{m}-\frac{i}{m^2} & 1 \leq i\leq m \\
2 & m+1 \leq i \leq n-m\\
1+\frac{n-i}{m+1} & n-m+1 \leq i \leq n \end{array} \right.,
$$
the $Y_{(i)}$'s are
$$\left \{\begin{array}{lc}
Y_{(i-m)}=a+\frac{i-1}{m}(X_{(1)}-a) & 1 \leq i\leq m \\
Y_{(i)}=X_{(i)} & m+1 \leq i \leq n-m\\
Y_{(i+m)}=b-\frac{n-i}{m}(b-X_{(n)}) & n-m+1 \leq i \leq n \end{array} \right.,
$$
and $a$ and $b$ are constants to be determined such that $P(a \leq X \leq b)\approx 1$. For example, when $F$ (the population distribution function) has a bounded support, $a$ and $b$ are lower and upper bound, respectively (for uniform(0,1) distribution, $a=0$ and $b=1$); if $F$ is bounded below (above), then $a (b)$ is lower (upper) support, $a=\bar{x}-k s \,(b=\bar{x}+k s)$, where
$$\bar{x}=\frac{1}{n}\sum_{i=1}^n x_i, \quad s^2=\frac{1}{n-1} \sum_{i=1}^n (x_i-\bar{x})^2,$$
and $k$ is a suitable number say 3 to 5 (for exponential distribution, $a=0$ and $b=\bar{x}+k s$); in the case that $F$ has no bound on its support, $a$ and $b$ may be chosen as $a=\bar{x}-k s$ and $b=\bar{x}+k s$.

Correa (1995) proposed another entropy estimator defined as
\begin{equation}\label{hc}
HC_{m,n}=-\frac{1}{n}\sum_{i=1}^n \log\left\{ \frac{\sum_{j=i-m}^{i+m} (X_{(j)}-\bar{X}_{(i)})(j-i)}{n \sum_{j=i-m}^{i+m} (X_{(j)}-\bar{X}_{(i)})^2}\right\},
\end{equation}
where
$$\bar{X}_{(i)}=\frac{1}{2m+1} \sum_{j=i-m}^{i+m} X_{(j)}.$$

Yousefzadeh and Arghami (2008) introduced the following entropy estimator
\begin{equation}\label{you}
HY_{m,n}=\sum_{i=1}^n \left\{ \frac{\hat{F}_y\left(X_{(i+m)} \right)-\hat{F}_y\left(X_{(i-m)} \right)}{\sum_{j=1}^n  \hat{F}_y\left(X_{(j+m)} \right)-\hat{F}_y\left(X_{(j-m)} \right) } \right\} \log\left\{\frac{ X_{(i+m)}-X_{(i-m)} }{ \hat{F}_y\left(X_{(i+m)} \right)-\hat{F}_y\left(X_{(i-m)} \right) }\right\},
\end{equation}
where for $i=2,\ldots,n-1$,
$$\hat{F}_y\left(X_{(i)}\right)=\frac{n-1}{n(n+1)} \left( i+\frac{1}{n-1}+ \frac{X_{(i)}-X_{(i-1)}}{X_{(i+1)}-X_{(i-1)}} \right),$$
and
$$\hat{F}_y\left(X_{(1)}\right)=1-\hat{F}_y\left(X_{(n)}\right)=\frac{1}{n+1}.$$

Alizadeh Noughabi (2010) developed an entropy estimator using kernel density estimator. His estimator is defined as
\begin{equation}\label{ali}
HA_{m,n}=-\frac{1}{n}\sum_{i=1}^n \log\left\{\frac{ \hat{f}\left(X_{(i+m)} \right)-\hat{f}\left(X_{(i-m)} \right) }{2}\right\},
\end{equation}
where $\hat{f}(x)$ is just as given in (\ref{bow}).\\
{\bf Remark 2:} In the all new entropy estimators which employ spacings of the order statistics, it is assumed that $m$ is an integer satisfying $1\leq m \leq n/2$, unless otherwise stated.

The test statistics obtained by replacing $HV_{m,n}$ in (\ref{Dinf}) with $HB_n$, $HVE_{m,n}$, $HE_{m,n}$, $HC_{m,n}$, $HY_{m,n}$ and $HA_{m,n}$ will be denoted by $\hat{D}_2$, $\hat{D}_3$, $\hat{D}_4$, $\hat{D}_5$, $\hat{D}_6$ and $\hat{D}_7$, respectively. Again, Monte Carlo approach is adopted to compute critical values of the resulting tests. To calculate test statistics based on the KL distance (with the exception of $\hat{D}_2$), the window size $m$ corresponding to a given sample size must be selected in advance. In entropy estimation based on spacings, choosing optimal $m$ for given $n$ is still an open problem. For each $n$, the window size having smallest critical value tends to yield greater power. For sample sizes 10, 20, 30 and 50, window sizes producing the minimum critical values for different tests are given in Table 1. Table 2 contains 0.05 critical points of the tests considered in this study. For the KL distance based tests, the above mentioned optimal window sizes are used. These thresholds will be used in the next section to study the power properties.

The entropy estimators mentioned in this section are consistent. In proving this result for the estimators dependent on the window size, it is assumed that $m/n\rightarrow 0$ as $m \rightarrow \infty$ and $n \rightarrow \infty$. See pertinent references for more details. The next proposition attends to optimal property of the tests based on the KL distance.\\
{\bf Proposition 1} {\it The tests based on $\hat{D}_i$, $i=1,\ldots,7$, are consistent.}\\
{\bf Proof.} Let $X_1,\ldots,X_n$ be a random sample of size $n$ from a population with density function $f_0(x;\mu,\sigma)$ given in (\ref{dens}). It is easy to see that for any $\mu \in \mathbb{R}$ and $\sigma>0$,
$$\frac{1}{n} \sum_{i=1}^n \log\left(f_0(X_i;\mu,\sigma) \right) \stackrel{a.s.}\rightarrow E\Big\{\log \left(f_0(X_i;\mu,\sigma) \right)\Big\}.$$
We may now conclude that
$$\frac{1}{n} \sum_{i=1}^n \log\left(f_0(X_i;\hat{\mu},\hat{\sigma}) \right) \stackrel{a.s.}\rightarrow E\Big\{\log \left(f_0(X_i;\mu,\sigma) \right)\Big\},$$
because $\hat{\mu}$ and $\hat{\sigma}$, defined in (\ref{mhat}) and (\ref{shat}), are strongly consistent estimators. Let $H_n$ be a typical entropy estimator. From consistency of $H_n$, we have
$$H_n\stackrel{p}\rightarrow - E\Big\{\log \left(f_0(X_i;\mu,\sigma) \right)\Big\}.$$
Putting these together, it follows that under the null hypothesis $\hat{D}_i \stackrel{p}\rightarrow 1$. Now, the result follows from the fact that $\hat{D}_i \stackrel{p}\rightarrow d'(>1)$ under the alternative hypothesis.\quad $\square$

\begin{table}
\begin{center}
\caption{The optimal window sizes for the tests of size 0.05 based on the KL distance}
\begin{tabular}{c c c c c c c} \hline
 &\multicolumn{6}{c}{Statistic}\\
\cline{2-7}
{$n$}&{$\hat{D}_1$}&{$\hat{D}_3$}&{$\hat{D}_4$}&{$\hat{D}_5$}&{$\hat{D}_6$}&{$\hat{D}_7$}\\
\hline
10 & 2 & 9 & 5 & 2 & 5 & 5 \\
20 & 4 & 19 & 10 & 4 & 10 & 10 \\
30 & 8 & 29 & 15 & 11 & 15 & 15 \\
50 & 20 & 49 & 25 & 23 & 25 & 25 \\
\hline
\end{tabular}
\end{center}
\end{table}

\begin{table}
\begin{center}
\caption{0.05 critical points of the tests}
\begin{tabular}{c c c c c} \hline
 &\multicolumn{4}{c}{$n$}\\
\cline{2-5}
{Statistic}&{10}&{20}&{30}&{50}\\
\hline
$KS$            & 0.270 & 0.196 & 0.163 & 0.128 \\
& & & & \\
$A^2$           & 0.919 & 0.983 & 1.026 & 1.037 \\
& & & & \\
$W^2$           & 0.129 & 0.138 & 0.140 & 0.141 \\
& & & & \\
$D_{n,\lambda}$ & 0.152 & 0.152 & 0.159 & 0.160 \\
& & & & \\
$Z_K$           & 1.890 & 2.508 & 2.881 & 3.231 \\
& & & & \\
$Z_A$           & 3.755 & 3.615 & 3.541 & 3.461 \\
& & & & \\
$Z_C$           &12.423 &15.940 &17.834 &19.787 \\
& & & & \\
$\hat{D}_1$     & 2.088 & 1.464 & 1.244 & 0.940 \\
& & & & \\
$\hat{D}_2$     & 1.274 & 1.158 & 1.103 & 1.042 \\
& & & & \\
$\hat{D}_3$     & 1.332 & 0.975 & 0.763 & 0.526 \\
& & & & \\
$\hat{D}_4$     & 0.842 & 0.740 & 0.653 & 0.531 \\
& & & & \\
$\hat{D}_5$     & 1.757 & 1.263 & 1.042 & 0.734 \\
& & & & \\
$\hat{D}_6$     & 1.367 & 1.109 & 0.924 & 0.689 \\
& & & & \\
$\hat{D}_7$     & 1.117 & 0.865 & 0.740 & 0.614 \\
\hline
\end{tabular}
\end{center}
\end{table}

\section{Power comparisons}
In this section, performances of the proposed tests are evaluated via Monte Carlo experiments. Toward this end, we considered nine families of alternatives:
\begin{itemize}
\item $t$ distribution with $n$ degrees of freedom denoted by $t_n$.

\item Normal distribution with mean $\mu$ and variance $\sigma^2$ denoted by N$(\mu,\sigma^2)$.

\item Logistic distribution with mean $\mu$ and variance $\pi^2 \sigma^2/3$ denoted by Lo$(\mu,\sigma^2)$.

\item Laplace distribution with mean $\mu$ and variance $2 \sigma^2$ denoted by La$(\mu,\sigma^2)$.

\item Gumbel distribution with mean $\mu+\sigma \gamma$ (where $\gamma$ is Euler's constant) and variance $\pi^2 \sigma^2/6$ denoted by Gu$(\mu,\sigma^2)$.

\item Beta distribution with mean $\alpha/(\alpha+\beta)$ denoted by Be$(\alpha,\beta)$.

\item Gamma distribution with mean $\alpha \beta$ and variance $\alpha \beta^2$ denoted by Ga$(\alpha,\beta)$.

\item Mixture of the normal and Cauchy distributions with mixing probability $p$ denoted by NC$(p,1-p)$. The distribution mixes N(0,1) and C(0,1) with weights $p$ and $1-p$, respectively.

\item Tukey distribution with parameter $h$ denoted by Tu$(h)$. It is distribution of the random variable $Z \exp\{Z h^2/2\}$ with $Z\sim N(0,1).$

\end{itemize}
The members selected from the above families are $t_3$, $t_5$, N(0,1), Lo(0,1), La(0,1), Gu(0,1), Be(2,1), Ga(2,1), NC(0.3,0.7) and Tu(1). For each alternative, 50,000 samples of sizes $n=10,20,30,50$ were generated, and the power of each test was estimated by the percentages of samples entering the rejection region. Tables 3-6 present the estimated powers of the fourteen tests of size 0.05 (given in Sections 2 and 3) for different sample sizes. To provide enough space for the outputs, the reference to parameters of the distributions is only made in the case of $t$ distribution. For each alternative, power entry associated with the best test among $\hat{D}_i$'s is in bold. In addition, the highest power value from the other tests is in italic.

{\linespread{1.6}
\begin{table}
\begin{center}
\caption{Power comparison for the tests of size 0.05 against several alternative distributions for $n=10$}
\begin{tabular}{c c c c c c c c c c c } \hline
&\multicolumn{10}{c}{Alternative}\\
\cline{2-11}
{Statistic}&{$t_3$}&{$t_5$}&{N}&{Lo}&{La}&{Gu}&{Be}&{Ga}&{NC}&{Tu}\\
\hline
KS              & \emph{0.028} & \emph{0.028} & 0.031 & \emph{0.028} & \emph{0.028} & \emph{0.048} & \emph{0.097} & \emph{0.086} & 0.040 & 0.061 \\
& & & & & & & & & & \\
$A^2$           & 0.013 & 0.013 & 0.016 & 0.012 & 0.013 & 0.024 & 0.051 & 0.044 & 0.037 & \emph{0.069} \\
& & & & & & & & & & \\
$W^2$           & \emph{0.028} & \emph{0.028} & \emph{0.033} & \emph{0.028} & \emph{0.028} & 0.047 & 0.085 & 0.076 & 0.040 & 0.063 \\
& & & & & & & & & & \\
$D_{n,\lambda}$ & 0.005 & 0.004 & 0.003 & 0.003 & 0.008 & 0.003 & 0.002 & 0.003 & 0.032 & 0.067 \\
& & & & & & & & & & \\
$Z_K$           & 0.012 & 0.012 & 0.013 & 0.011 & 0.012 & 0.024 & 0.057 & 0.048 & 0.041 & 0.068 \\
& & & & & & & & & & \\
$Z_A$           & 0.016 & 0.016 & 0.021 & 0.016 & 0.016 & 0.033 & 0.079 & 0.060 & 0.041 & 0.067 \\
& & & & & & & & & & \\
$Z_C$           & 0.008 & 0.010 & 0.014 & 0.010 & 0.010 & 0.020 & 0.054 & 0.036 & \emph{0.042} & 0.068 \\
& & & & & & & & & & \\
$\hat{D}_1$     & 0.114 & 0.145 & 0.202 & 0.159 & 0.103 & 0.210 & 0.423 & 0.282 & 0.061 & \textbf{0.046} \\
& & & & & & & & & & \\
$\hat{D}_2$     & 0.177 & 0.230 & 0.329 & 0.257 & 0.158 & 0.295 & 0.502 & 0.322 & 0.074 & 0.038 \\
& & & & & & & & & & \\
$\hat{D}_3$     & 0.178 & 0.232 & 0.330 & 0.260 & 0.162 & 0.296 & \textbf{0.515} & 0.326 & 0.074 & 0.038 \\
& & & & & & & & & & \\
$\hat{D}_4$     & \textbf{0.192} & \textbf{0.252} & \textbf{0.356} & \textbf{0.280} & \textbf{0.172} & \textbf{0.304} & 0.496 & 0.311 & \textbf{0.077} & 0.037 \\
& & & & & & & & & & \\
$\hat{D}_5$     & 0.123 & 0.157 & 0.220 & 0.173 & 0.110 & 0.225 & 0.441 & 0.298 & 0.063 & 0.045 \\
& & & & & & & & & & \\
$\hat{D}_6$     & 0.168 & 0.219 & 0.312 & 0.245 & 0.153 & 0.286 & 0.512 & 0.325 & 0.072 & 0.038 \\
& & & & & & & & & & \\
$\hat{D}_7$     & 0.156 & 0.201 & 0.287 & 0.224 & 0.142 & 0.277 & 0.506 & \textbf{0.335} & 0.069 & 0.039 \\
\hline
\end{tabular}
\end{center}
\end{table}

\begin{table}
\begin{center}
\caption{Power comparison for the tests of size 0.05 against several alternative distributions for $n=20$}
\begin{tabular}{c c c c c c c c c c c } \hline
&\multicolumn{10}{c}{Alternative}\\
\cline{2-11}
{Statistic}&{$t_3$}&{$t_5$}&{N}&{Lo}&{La}&{Gu}&{Be}&{Ga}&{NC}&{Tu}\\
\hline
KS              & 0.042 & 0.049 & 0.063 & 0.052 & 0.040 & 0.127 & 0.343 & 0.279 & 0.043 & 0.060 \\
& & & & & & & & & & \\
$A^2$           & 0.028 & 0.036 & 0.059 & 0.042 & 0.027 & 0.095 & 0.247 & 0.183 & 0.036 & 0.072 \\
& & & & & & & & & & \\
$W^2$           & 0.039 & 0.048 & 0.065 & 0.052 & 0.038 & 0.105 & 0.231 & 0.192 & 0.039 & 0.061 \\
& & & & & & & & & & \\
$D_{n,\lambda}$ & 0.035 & 0.059 & 0.122 & 0.073 & 0.031 & 0.126 & 0.365 & 0.184 & 0.030 & \emph{0.078} \\
& & & & & & & & & & \\
$Z_K$           & 0.030 & 0.038 & 0.059 & 0.042 & 0.026 & 0.137 & 0.417 & 0.333 & 0.041 & 0.070 \\
& & & & & & & & & & \\
$Z_A$           & \emph{0.094} & \emph{0.140} & \emph{0.261} & \emph{0.167} & \emph{0.076} & \emph{0.313} & \emph{0.688} & \emph{0.492} & \emph{0.053} & 0.059 \\
& & & & & & & & & & \\
$Z_C$           & 0.061 & 0.098 & 0.195 & 0.118 & 0.049 & 0.218 & 0.565 & 0.343 & 0.047 & 0.068 \\
& & & & & & & & & & \\
$\hat{D}_1$     & 0.354 & 0.501 & 0.739 & 0.573 & 0.334 & 0.733 & 0.974 & 0.852 & 0.084 & \textbf{0.036} \\
& & & & & & & & & & \\
$\hat{D}_2$     & 0.441 & 0.606 & 0.812 & 0.675 & 0.404 & \textbf{0.765} & 0.965 & 0.798 & \textbf{0.098} & 0.030 \\
& & & & & & & & & & \\
$\hat{D}_3$     & 0.416 & 0.592 & 0.840 & 0.678 & 0.428 & 0.716 & 0.968 & 0.717 & 0.086 & 0.032 \\
& & & & & & & & & & \\
$\hat{D}_4$     & \textbf{0.453} & \textbf{0.639} & \textbf{0.873} & \textbf{0.728} & \textbf{0.456} & 0.711 & 0.947 & 0.669 & 0.091 & 0.031 \\
& & & & & & & & & & \\
$\hat{D}_5$     & 0.365 & 0.520 & 0.761 & 0.593 & 0.351 & 0.745 & 0.975 & 0.852 & 0.085 & 0.035 \\
& & & & & & & & & & \\
$\hat{D}_6$     & 0.410 & 0.581 & 0.826 & 0.666 & 0.416 & 0.742 & \textbf{0.977} & 0.779 & 0.086 & 0.032 \\
& & & & & & & & & & \\
$\hat{D}_7$     & 0.301 & 0.412 & 0.611 & 0.465 & 0.274 & 0.712 & 0.966 & \textbf{0.893} & 0.084 & 0.035 \\
\hline
\end{tabular}
\end{center}
\end{table}

\begin{table}
\begin{center}
\caption{Power comparison for the tests of size 0.05 against several alternative distributions for $n=30$}
\begin{tabular}{c c c c c c c c c c c } \hline
&\multicolumn{10}{c}{Alternative}\\
\cline{2-11}
{Statistic}&{$t_3$}&{$t_5$}&{N}&{Lo}&{La}&{Gu}&{Be}&{Ga}&{NC}&{Tu}\\
\hline
KS              & 0.058 & 0.071 & 0.106 & 0.078 & 0.046 & 0.247 & 0.661 & 0.546 & 0.048 & 0.060 \\
& & & & & & & & & & \\
$A^2$           & 0.050 & 0.077 & 0.146 & 0.092 & 0.040 & 0.224 & 0.560 & 0.401 & 0.036 & 0.075 \\
& & & & & & & & & & \\
$W^2$           & 0.055 & 0.072 & 0.113 & 0.082 & 0.047 & 0.191 & 0.445 & 0.348 & 0.043 & 0.061 \\
& & & & & & & & & & \\
$D_{n,\lambda}$ & 0.123 & 0.225 & 0.455 & 0.283 & 0.100 & 0.417 & 0.811 & 0.519 & 0.031 & \emph{0.087} \\
& & & & & & & & & & \\
$Z_K$           & 0.066 & 0.099 & 0.189 & 0.116 & 0.047 & 0.417 & 0.864 & 0.776 & 0.046 & 0.073 \\
& & & & & & & & & & \\
$Z_A$           & \emph{0.245} & \emph{0.392} & \emph{0.669} & \emph{0.474} & \emph{0.203} & \emph{0.731} & \emph{0.974} & \emph{0.897} & \emph{0.065} & 0.054 \\
& & & & & & & & & & \\
$Z_C$           & 0.172 & 0.300 & 0.564 & 0.371 & 0.141 & 0.587 & 0.933 & 0.764 & 0.050 & 0.068 \\
& & & & & & & & & & \\
$\hat{D}_1$     & 0.584 & 0.791 & 0.974 & 0.880 & 0.633 & 0.962 & \textbf{1} & 0.988 & 0.086 & 0.030 \\
& & & & & & & & & & \\
$\hat{D}_2$     & \textbf{0.674} & 0.855 & 0.975 & 0.914 & 0.659 & 0.960 & \textbf{1} & 0.964 & \textbf{0.105} & 0.026 \\
& & & & & & & & & & \\
$\hat{D}_3$     & 0.602 & 0.811 & 0.986 & 0.906 & 0.690 & 0.918 & \textbf{1} & 0.907 & 0.080 & 0.030 \\
& & & & & & & & & & \\
$\hat{D}_4$     & 0.655 & \textbf{0.861} & \textbf{0.993} & \textbf{0.941} & \textbf{0.738} & 0.920 & 0.999 & 0.887 & 0.087 & 0.029 \\
& & & & & & & & & & \\
$\hat{D}_5$     & 0.614 & 0.819 & 0.983 & 0.905 & 0.670 & \textbf{0.965} & \textbf{1} & 0.983 & 0.088 & 0.029 \\
& & & & & & & & & & \\
$\hat{D}_6$     & 0.604 & 0.811 & 0.983 & 0.901 & 0.678 & 0.947 & \textbf{1} & 0.963 & 0.083 & 0.030 \\
& & & & & & & & & & \\
$\hat{D}_7$     & 0.434 & 0.596 & 0.841 & 0.673 & 0.412 & 0.935 & \textbf{1} & \textbf{0.997} & 0.087 & \textbf{0.031} \\
\hline
\end{tabular}
\end{center}
\end{table}

\begin{table}
\begin{center}
\caption{Power comparison for the tests of size 0.05 against several alternative distributions for $n=50$}
\begin{tabular}{c c c c c c c c c c c } \hline
&\multicolumn{10}{c}{Alternative}\\
\cline{2-11}
{Statistic}&{$t_3$}&{$t_5$}&{N}&{Lo}&{La}&{Gu}&{Be}&{Ga}&{NC}&{Tu}\\
\hline
KS              & 0.095 & 0.137 & 0.253 & 0.151 & 0.063 & 0.583 & 0.976 & 0.928 & 0.054 & 0.058 \\
& & & & & & & & & & \\
$A^2$           & 0.142 & 0.261 & 0.517 & 0.316 & 0.099 & 0.619 & 0.955 & 0.847 & 0.040 & 0.079 \\
& & & & & & & & & & \\
$W^2$           & 0.096 & 0.148 & 0.281 & 0.169 & 0.069 & 0.421 & 0.828 & 0.689 & 0.047 & 0.064 \\
& & & & & & & & & & \\
$D_{n,\lambda}$ & 0.400 & 0.644 & 0.906 & 0.740 & 0.328 & 0.874 & 0.997 & 0.937 & 0.043 & \emph{0.099} \\
& & & & & & & & & & \\
$Z_K$           & 0.228 & 0.385 & 0.701 & 0.462 & 0.168 & 0.933 & \emph{1} & 0.998 & 0.058 & 0.075 \\
& & & & & & & & & & \\
$Z_A$           & \emph{0.622} & \emph{0.853} & \emph{0.988} & \emph{0.924} & \emph{0.603} & \emph{0.995} & \emph{1} & \emph{1} & \emph{0.088} & 0.051 \\
& & & & & & & & & & \\
$Z_C$           & 0.514 & 0.774 & 0.970 & 0.866 & 0.482 & 0.974 & \emph{1} & 0.996 & 0.063 & 0.069 \\
& & & & & & & & & & \\
$\hat{D}_1$     & 0.815 & 0.965 & \textbf{1} & 0.996 & 0.939 & \textbf{1} & \textbf{1} & \textbf{1} & 0.083 & 0.029 \\
& & & & & & & & & & \\
$\hat{D}_2$     & \textbf{0.917} & \textbf{0.992} & \textbf{1} & 0.998 & 0.947 & \textbf{1} & \textbf{1} & \textbf{1} & \textbf{0.115} & 0.022 \\
& & & & & & & & & & \\
$\hat{D}_3$     & 0.803 & 0.960 & \textbf{1} & 0.996 & 0.949 & 0.995 & \textbf{1} & 0.993 & 0.078 & \textbf{0.030} \\
& & & & & & & & & & \\
$\hat{D}_4$     & 0.852 & 0.979 & \textbf{1} & \textbf{0.999} & \textbf{0.970} & 0.997 & \textbf{1} & 0.995 & 0.086 & 0.028 \\
& & & & & & & & & & \\
$\hat{D}_5$     & 0.820 & 0.967 & \textbf{1} & 0.997 & 0.945 & \textbf{1} & \textbf{1} & \textbf{1} & 0.083 & 0.029 \\
& & & & & & & & & & \\
$\hat{D}_6$     & 0.813 & 0.964 & \textbf{1} & 0.996 & 0.943 & 0.999 & \textbf{1} & \textbf{1} & 0.080 & 0.029 \\
& & & & & & & & & & \\
$\hat{D}_7$     & 0.786 & 0.943 & 0.999 & 0.979 & 0.827 & \textbf{1} & \textbf{1} & \textbf{1} & 0.094 & 0.027 \\
\hline
\end{tabular}
\end{center}
\end{table}

\begin{table}
\begin{center}
\caption{Power differences between the best test among $\hat{D}_i$'s and the best of other tests}
\begin{tabular}{c c c c c c c c c c c } \hline
&\multicolumn{10}{c}{Alternative}\\
\cline{2-11}
{$n$}&{$t_3$}&{$t_5$}&{N}&{Lo}&{La}&{Gu}&{Be}&{Ga}&{NC}&{Tu}\\
\hline
10 & 0.164 & 0.224 & 0.323 & 0.252 & 0.144 & 0.256 & 0.418 & 0.249 & 0.035 & -0.023 \\
& & & & & & & & & & \\
20 & 0.359 & 0.499 & 0.612 & 0.561 & 0.380 & 0.452 & 0.289 & 0.401 & 0.045 & -0.042 \\
& & & & & & & & & & \\
30 & 0.429 & 0.469 & 0.324 & 0.467 & 0.535 & 0.234 & 0.026 & 0.100 & 0.040 & -0.056 \\
& & & & & & & & & & \\
50 & 0.295 & 0.139 & 0.012 & 0.075 & 0.367 & 0.005 & 0 & 0 & 0.027 & -0.069 \\
\hline
\end{tabular}
\end{center}
\end{table}
}

It is observed that no single test is uniformly most powerful. We note, however, that the tests based on the KL distance are generally more powerful than the other tests. Compare the bold and italic entries for each alternative. Given a distribution and sample size, difference of the italic entry from the bold one is reported in Table 7. The values are sizable for symmetric distributions like $t_3$, $t_5$, N(0,1), Lo(0,1), La(0,1) and Gu(0,1). All of the tests perform poorly when the parent distribution is either NC(0.3,0.7) or Tu(1), and increasing the sample size does not give rise to marked improvement in power.

With the exception of sample size 10, $Z_A$ is generally the best among KS, $A^2$, $W^2$, $D_{n,\lambda}$, $Z_K$, $Z_A$ and $Z_C$ tests. Moreover, it can be seen that either $\hat{D}_2$ or $\hat{D}_4$ has mostly the best performance among $\hat{D}_i$'s.

\section{Example}
Heavy-tailed distributions, like Cauchy, are better models for financial returns because the normal model does not capture the large fluctuations seen in real assets. Nolan (2014) provides an accessible introduction to financial modeling using such distributions.

The stock market return is the return that we obtain from stock market by buying and selling stocks or get dividends by the company whose stock you hold. The stock market price is usually modeled by lognormal distribution, that is to say stock market returns follow the Gaussian law. The feature of stock market return distribution is a sharp peak and heavy tails. The Gaussian distribution clearly does not enjoy these attributes. So the Cauchy distribution may be a potential model. We now apply the fourteen goodness-of-fit tests to 30 returns of closing prices of the German Stock Index (DAX). The data are observed daily from January 1, 1991, excluding weekends and public holidays. The data (rounded up to seven decimal places) are given in Table 8. The Cauchy Q-Q plot appears in Figure 1. The corresponding histogram, superimposed by a Cauchy density function, is also included. The location and scale parameters estimated from the data are $\hat{\mu}=0.0009629174$ and $\hat{\sigma}=0.003635871$.

\begin{figure}[h!]
\centering
\includegraphics[]{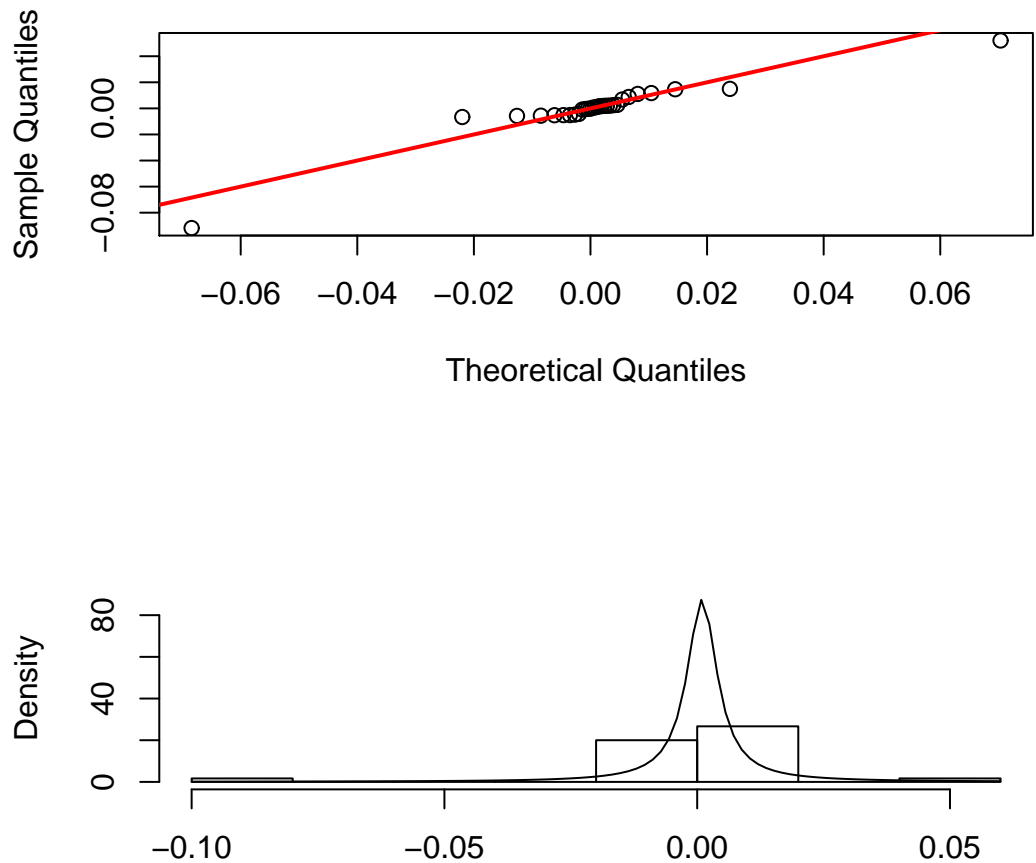}  
\vspace{-1cm}
\caption{The Cauchy Q-Q plot of the 30 returns, and the corresponding histogram along with fitted Cauchy density}
\end{figure}

\begin{table}
\begin{center}
\caption{Scores for 30 returns of closing prices of DAX }
\begin{tabular}{ c c c c c c } \hline
0.0011848 & -0.0057591 & -0.0051393 & -0.0051781 &  0.0020043  & 0.0017787 \\
0.0026787 & -0.0066238 & -0.0047866 & -0.0052497 &  0.0004985  & 0.0068006 \\
0.0016206 & 0.0007411  & -0.0005060 &  0.0020992 & -0.0056005  & 0.0110844 \\
-0.0009192& 0.0019014  & -0.0042364 &  0.0146814 & -0.0002242  & 0.0024545 \\
-0.0003083& -0.0917876 & 0.0149552  & 0.0520705  &  0.0117482  & 0.0087458 \\
\hline
\end{tabular}
\end{center}
\end{table}

The values of all statistics are computed (see Table 9), and compared with the corresponding critical values in Table 2. By using any test, the null hypothesis that the data follow the Cauchy distribution is not rejected at 0.05 significance level.

\begin{table}
\centering
\caption{Observed values of the different statistics}
\begin{tabular}{c c c c c c c}
\hline
{KS} & {$A^2$} & {$W^2$} & {$D_{n,\lambda}$} & {$Z_K$} & {$Z_A$} & {$Z_C$}\\
 0.126 & 0.498 & 0.076 & 0.051 & 1.343 & 3.346 & 5.761 \\
\hline \hline
{$\hat{D}_1$} & {$\hat{D}_2$} & {$\hat{D}_3$} & {$\hat{D}_4$} & {$\hat{D}_5$} & {$\hat{D}_6$} & {$\hat{D}_7$}\\
 0.661 & 0.844 & 0.255 & 0.302 & 0.386 & 0.358 & 0.461 \\
\hline
\end{tabular}
\end{table}

\section{Conclusion}
A number of theoretical and empirical studies suggest that the financial data are mostly modeled using densities with sharp peak and fat tails. The Gaussian distribution, however, cannot account for the excess kurtosis effectively. The Cauchy distribution thus appears to be a viable alternate. For example, it is a good model for financial returns because of the power to capture the large fluctuations seen in real assets.

This article concerns goodness-of-fit test for the Cauchy distribution. Six tests based on the KL information criterion are developed, and shown to be consistent. A simulation study is carried out to compare the performances of the new tests with their contenders. In doing so, five sample sizes and nine families of alternatives are considered. It emerges that the new tests are powerful against many symmetric distributions. The proposed procedures are finally applied on real data example.

The authors plan to develop similar tests using more recent entropy estimators, especially those which are based on numerical methods rather than spacings. To overcome the problem of window size selection, one may follow a novel approach suggested by Vexler and Gurevich (2010). The use of maximum likelihood estimators and minimum discriminant information loss estimators also deserve investigation. We hope to report our findings in the future.

\section*{References}
\begin{description}
\item Alizadeh Noughabi, H. (2010). A new estimator of entropy and its application in testing normality. Journal of Statistical Computation and Simulation 80, 1151-1162.

\item Bowman, A.W. (1992). Density based tests for goodness-of-fit. Journal of Statistical Computation and Simulation 40, 1-13.

\item Choi, B., and Kim, K. (2006). Testing goodness-of-fit for laplace distribution based on maximum entropy. Statistics 40, 517-531.

\item Correa, J.C. (1995). A new estimator of entropy. Communications in Statistics: Theory and Methods 24, 2439-2449.

\item Dudewicz, E.J., and van der Meulen, E.C. (1981). Entropy-based tests of uniformity. Journal of the American Statistical Association 76, 967-974.

\item Ebrahimi, N., Pflughoeft, K., and Soofi, E.S. (1994). Two measures of sample entropy. Statistics and Probability Letters 20, 225-234.

\item Grzegorzewski, P., and Wieczorkowski, R. (1999). Entropy based goodness-of-fit test for exponentiality. Communications in Statistics: Theory and Methods 28, 1183-1202.

\item G\"{u}rtler, N., and Henze, N. (2000). Goodness-of-fit tests for the Cauchy distribution based on the
empirical characteristic function. Annals of the Institute of Statistical Mathematics 52, 267-286.

\item Johnson, N.L., Kotz, S., and Balakrishnan, N. (1994). Continuous Univariate Distributions, Volume 1, 2nd Edition. Wiley, New York.

\item Kagan, Y.Y. (1992). Correlations of earthquake focal mechanisms. Geophysical Journal International 110, 305-320.

\item Kullback, S. (1997). Information Theory and Statistics. Dover Publications, New York.

\item Mahdizadeh, M., and Zamanzade, E. (2016). New goodness-of-fit tests for the Cauchy distribution. Journal of Applied Statistics, DOI: 10.1080/02664763.2016.1193726

\item Min, I.A., Mezic, I., and Leonard, A. (1996). Levy stable distributions for velocity and velocity difference in systems of vortex elements. Physics of Fluids 8, 1169-1180.

\item Mudholkar, G.S., and Tian, L. (2002). An entropy characterization of the inverse Gaussian distribution and related goodness-of-fit test. Journal of Statistical Planning and Inference 102, 211-221.

\item Nolan, J.P. (2014). Financial modeling with heavy-tailed stable distributions. WIREs Computational Statistics 6, 45-55.

\item Roe, B.P. (1992). Probability and Statistics in Experimental Physics. Springer, New York.

\item Stapf, S., Kimmich, R., Seitter, R.O., Maklakov, A.I., and Skid, V.D. (1996). Proton and deuteron field-cycling NMR relaxometry of liquids confined in porous glasses. Colloids and Surfaces: A Physicochemical and Engineering Aspects 115, 107-114.

\item Stigler, S.M. (1974). Cauchy and the witch of Agnesi: an historical note on the Cauchy distribution. Biometrika 61, 375-380.

\item Van Es, B. (1992). Estimating functionals related to a density by class of statistics based on spacings. Scandinavian Journal of Statistics 19, 61-72.

\item Vasicek, O. (1976). A test of normality based on sample entropy. Journal of the Royal Statistical Society B 38, 54-59.

\item Vexler, A. and Gurevich, G. (2010). Empirical likelihood ratios applied to goodness-of-fit tests based on
sample entropy. Comput. Stat. Data Anal. 54, 531-545.

\item Winterton, S.S., Smy, T.J., and Tarr, N.G. (1992). On the source of scatter in contact resistance data. Journal of Electronic Materials 21, 917-921.

\item Yousefzadeh, F., and Arghami, N.R. (2008). Testing exponentiality based on type II censored data and a new CDF estimator. Communications in Statistics: Simulation and Computation 37, 1479-1499.

\item Zhang, J. (2002). Powerful goodness-of-fit tests based on the likelihood ratio. Journal of the Royal Statistical Society B 64, 281-294.
\end{description}

\end{document}